\magnification=1200 
\parindent=3em 
\baselineskip=24truept 
\parskip=0pt 
\overfullrule=0pt

\def\sqr#1#2{{\vcenter{\hrule height.#2pt 
    \hbox{\vrule width.#2pt height#1pt \kern#1pt 
      \vrule width .#2pt} 
    \hrule height.#2pt}}}

\def\ts{\textstyle}
 
\def\ab{{\alpha\beta}} 
 
\def\dd#1#2{{d#1\over d#2}} 
\def\pp#1#2{{\partial #1\over\partial #2}} 
%
%

\def\oot{{1\ov 2}} 
 
\def\cl{\centerline}

\def\l{\left} 
\def\r{\right} 
\def\cd{\cdot}

\def\sq{\sqrt}

\def\mn{{\mu\nu}} 
\def\nm{{\nu\mu}}

\def\Ga{\Gamma}

\def\al{\alpha} 
\def\be{\beta}

\def\la{\lambda} 
\def\ga{\gamma} 
\def\de{\delta} 
\def\pa{\partial} 
\def\inf{\infty} 
 
\def\vp{\varphi}

\def\th{\theta} 
\def\ep{\epsilon} 
 
\def\ka{\kappa}

\def\ov{\over} 
 
\def\ra{\rightarrow} 
\def\Ra{\Rightarrow} 
 
\def\const{{\rm const.}} 
 
\def\ln{{\rm ln}} 
\def\gapprox{\lower.4ex\hbox{$\;\buildrel >\over{\scriptstyle\sim}\;$}} 
\def\lapprox{\lower.4ex\hbox{$\;\buildrel <\over{\scriptstyle\sim}\;$}}

%

\font\tenbi=cmbxti10 
\newfam\bifam \def\bi{\fam\bifam\tenbi} \textfont\bifam=\tenbi 
 
\font\tensi=cmsy10 
\newfam\sifam  \textfont\sifam=\tensi 
\font\squinttenbi=cmbxti10 at 9pt 
\scriptfont\bifam=\squinttenbi

\def\vecA{{\bi{A}}} 
\def\vecB{{\bi{B}}}

\def\vecE{{\bi{E}}}

\def\vecL{{\bi{L}}}

\def\fpl{{\it Found. Phys. Lett.} }

\def\pla{{\it  Phys. Lett. A} }

\def\prd{{\it Phys. Rev. D} }

\parindent=1.65em 
\def\no#1{\item{#1.}}

\def\head#1{\bigskip\noindent{\bf #1}}

\cl{\bf THE SPEED OF LIGHT AS A DILATON FIELD} 
 
\bigskip 
\bigskip 
\cl{Walter Wyss} 
 
\bigskip 
\cl{\it Department of Physics} 
 
\cl{\it University of Colorado} 
 
\cl{\it Boulder, CO 80309} 
 
\vskip.5truein

Through dimensional analysis, eliminating the physical time, 
we identify the speed of light as a dilaton field.  This leads to a  
restmass zero, spin zero gauge field which we call the 
speedon field.  The complete Lagrangian for gravitational, 
electromagnetic and speedon field interactions with a 
charged scalar field, representing matter, is given.  We 
then find solutions for the gravitational-electromagnetic-speedon field 
equations.  This then gives an expression for the speed of light. 
 
\vfil\eject 
 
\head{I. Introduction} 
 
We are interested in the concept of physical time.  Physical 
time is defined with the help of a periodical physical system, 
e.g., an atomic clock.  At present we have adopted a universal 
time currency.  But what about local time currencies?:  commonly 
one uses the gravitational redshift as an exchange rate; an atomic 
clock in Boulder has a higher frequency than a similar 
atomic clock in Paris.  Are there other contributions to this 
exchange rate?  To look for an answer we  formulated [1] 
electrodynamics in such a way that the physical time 
never occurs.  Only the geometric time $x^o$, which 
is related to the physical time $t$ by $x^o= ct$ occurs. 
$c$ is the speed of light and knowing the geometric time 
one can recover the physical time.  In Special Relativity and 
also in Einstein's Theory of Gravity only the geometric time 
occurs; it has the physical dimension of a length.  The speed 
of light only enters if one wants to convert to physical time.  
The speed of light thus does not have to be a 
constant, but is related to the concept of physical time.  Indeed, 
in the formulation of electrodynamics, written independently of  
physical time, the speed of light enters as a scale factor.  It can thus 
be interpreted as a dilaton field [2].  Since this field is related to the  
speed of light we call it the speedon field.  It belongs to the 
restmass zero, spin zero representation of the Lorentz group. 
The corresponding elementary agent we called the ``speedon."  We 
thus have a trinity of restmass zero gauge particles:  speedon 
(spin zero), photon (spin one) and graviton (spin two). 
 
For interactions with  a charged scalar field representing matter  
each of these gauge fields contribute through their own covariant derivative. 
 
In Chapter II we give the general formulation for gravitational interactions 
as described by a Lagrange variational principle. 
 
In Chapter III we present the complete interaction between the 
gravitational field, the electromagnetic field, the speedon field and 
a charged scalar field representing matter. 
 
In Chapter IV we solve the equations of motion in the absence of the 
matter field.  This gives a solution for the gravitational field 
and the electromagnetic field in terms of the speedon field.   
Turning off the  electromagnetic field we find a expression for 
the pure speedon field and for the corresponding speed of light. 
 
Appendix A describes the scaling of the electromagnetic field and the 
motivation to introduce the speed of light as a dilaton field. 
 
In Appendix C the gravitational interaction with a restmass scalar field 
is revisited. 
 
\head{II. Gravitational Interactions} 
 
With the notation in [3] gravitational interactions are given 
through a variational principle where the action is given by 
$$ 
A = \int dx \sq{g} R - 2 \ka  
\int dx \sq{g} L (g, \phi) = \int dxL 
\eqno({\rm II}.1) 
$$ 
and where $dx$ is the 4-volume element, $L(g,\phi)$ 
is the nongravitational Lagrangian, with $\phi$ any multicomponent 
field, and 
$$ 
\ka = { 8 \pi G \ov c^2}  
\eqno({\rm II}.2) 
$$ 
with $c$ being the speed of light and $G$ the  
universal gravitational constant. 
 
Observe that 
$$ 
G_o = {G \ov c^2} 
\eqno({\rm II}.3) 
$$ 
has the physical dimension 
$$ 
[G_o ] = M^{-1} L. 
\eqno({\rm II}.4) 
$$ 
Throughout, $M = $ mass, $L$ = length, $T=$ physical time, 
$Q=$ charge. 
 
Thus $G_o$ does not depend on the physical time, i.e., the 
concept of a second, and 
$$ 
\ka = 8 \pi G_o 
\eqno({\rm II}.5) 
$$ 
The action $A$ with physical dimension 
$$ 
[A]= L^2 
\eqno({\rm II}.6) 
$$ 
also does not depend on physical time.  The physical 
dimension of the nongravitational Lagrangian $L(g,\phi)$ is 
$$ 
[L(g,\phi)] = ML^{-3}, 
\eqno({\rm II}.7) 
$$ 
i.e., a mass density. 
 
For the Euler derivative with respect to the gravitational field $g_\ab$ 
we get 
$$ 
\ep(g_\ab) = {\pa \ov \pa g_\ab}- \pa_\mu  
{\pa \ov  \pa g_{\ab,\mu}} 
\eqno({\rm II}.8) 
$$ 
$$ 
\ep(g_\ab) [\sq{g} R] 
= -\sq{g} [ R^\ab - \oot g^\ab R] 
\eqno({\rm II}.9) 
$$ 
and  
$$ 
\ep(g_\ab) [ \sq{g}  L (g,\phi)] \equiv M^\ab, 
\eqno({\rm II}.10) 
$$ 
where $M^\ab$ is the so-called gravitational stress tensor. 
 
Introducing $T^\ab$ through 
$$ 
M^\ab \equiv - \oot \sq{g} T^\ab 
\eqno({\rm II}.11) 
$$ 
we then get Einstein's equation 
$$ 
R^\ab - \oot g^\ab R = \ka T^\ab 
\eqno({\rm II}.12) 
$$ 
Observe that 
$$ 
T^\ab = - 2 \ep (g_\ab) 
[L(g,\phi)] - g^\ab L (g,\phi) + 2 \Ga^\nu_{\ \, \mn} 
\pp  {L(g,\phi)} {g_{\ab,\mu}} 
\eqno({\rm II}.13) 
$$ 
In addition we have the equations of motion for 
the field $\phi$ 
$$ 
\ep (\phi) [\sq{g} L (g,\phi)] = 0 
\eqno({\rm II}.14) 
$$ 
Comments 
 
{\parindent=1.65em 
 
\item{1)} The action for gravitational interaction is invariant 
under the inhomogeneous Lorentz group.  In addition it is  
invariant under the gravitational gauge group [4] and any 
other gauge group or internal symmetry group. 
 
\item{2)}  The right-hand side of Einstein's equation (II.12) 
is not arbitrary but is related to $M^\ab$, which is an 
Euler derivative.  If one looks at the matter Lagrangian 
$L(\eta, \phi)$, i.e., the Lagrangian $L(g, \phi)$ where 
the gravitational field $g$ is replaced by the Minkowski metric 
$\eta$, then there is the concept of an energy-momentum tensor 
[5].  The right-hand side $T^\ab$ in Einstein's equation,  
evaluated for $g=\eta$ is exactly equal to the energy-momentum 
tensor belonging to $L(\eta, \phi)$ [6].  That this is true for any 
field is highly nontrivial [7] and is a consequence from the 
fact that the energy-momentum tensor for any gravitational 
interaction vanishes identically. 
 
\item{3)}  For dust one usually takes 
$$ 
T^\ab=\varrho u^\al u^\be, \ \ u^\al u_\al =1 
\eqno({\rm II}.15) 
$$ 
This, however, does not fit in the above scheme; 
the closest we can get to such a case is by looking at 
a massless scalar field, i.e., a dilaton field.  This field 
then also has an equation of motion.  For mathematical 
consistency the right-hand side of Einstein's equation should be an Euler 
derivative. 
 
} 
 
\head{III. Graviton, Photon, Speedon and a Charged Scalar Field} 
 
Here we give the most general theory of the interactions between 
a charged scalar field and gauge fields belonging to the rest 
mass zero [8].  This involves the restmass zero fields of spin 2 
(gravity), spin one (electrodynamics) and spin zero (the speed 
of light).  These are all gauge fields and bring their own covariant  
derivative.  The restmass zero, spin zero field, which is a dilaton 
field, is related to the scaling  of the electromagnetic field such 
that the physical time never enters the corresponding  
Lagrangian (Appendix A). 
 
We thus have the fields $g_\ab, L_\al,  S, \phi^+, \phi$ with 
the physical dimensions 
$$ 
[g_\ab] = 0, \ \ [L_\al] = MQ^{-1}, \ \  
[S] = 0, \ \ [\phi^+] = [\phi] = 0 
\eqno({\rm III}.1) 
$$ 
The action is given by 
$$ 
A=\int dxL 
\eqno({\rm III}.2) 
$$ 
where  
$$ 
L=\sq{g}R - 2 \la_o \sq{g} L_o (L_\al) - 2\sq{g}  
L_o (S) - 2 \sq{g} L_M (\phi). 
\eqno({\rm III}.3) 
$$ 
The coupling constant $\la_o$ is 
$$ 
\la_o = {2G_o \ov K_o} \quad , \quad  
K_o = {K \ov c^2} 
\eqno({\rm III}.4) 
$$ 
with $K$ being the Coulomb constant. 
$K_o$ and $\la_o$ have the physical dimensions 
$$ 
[K_o] = MLQ^{-2} \quad , \quad  
[\la_o] = M^{-2} Q^2 
\eqno({\rm III}.5) 
$$ 
All these constants do not depend on physical time. 
 
In what follows all indices are raised with the inverse 
gravitational field $g^\ab$ and lowered with $g_\ab$. 
The Lagrangian for the electromagnetic field $\{L_\al\}$ 
is given by 
$$ 
\eqalignno{ 
L_o (L_\al) &= {1 \ov 4} F_\mn F^\nm &({\rm III}.6)\cr 
F_\mn &\equiv \pa_\mu L_\nu - \pa_\nu L_\mu &({\rm III}.7)\cr 
} 
$$ 
The Lagrangian for the speedon field is 
$$ 
L_o (S) = (\pa_\mu S) (\pa^\mu S), 
\eqno({\rm III}.8) 
$$ 
and the Lagrangian for the matterfield is 
$$ 
L_M (\phi) = (\hat D_\mu \phi)^+ (\hat D^\mu \phi) 
- m^2 \phi^+ \phi 
\eqno({\rm III}.9) 
$$ 
where $\hat D_\mu$ is the overall covariant derivative given by 
$$ 
\eqalignno{ 
\hat D_\mu \phi &= D_\mu \phi + i \la L_\mu \phi 
+ i  (\pa_\mu S) \phi &({\rm III}.10)\cr 
(\hat D_\mu \phi)^+ &= D_\mu \phi^+ - i \la L_\mu 
\phi^+ - i (\pa_\mu S) \phi^+ &({\rm III}.11)\cr 
} 
$$ 
with $D_\mu$ being the geometric covariant derivative, 
belonging to the graviton.   
 
We also have the following physical dimensions of the coupling 
constants $m$ and $\la$, 
$$ 
[m] = L^{-1} \quad , \quad  
[\la] = M^{-1} L^{-1} Q 
\eqno({\rm III}.12) 
$$ 
 
We now get the equations of motion 
 
\noindent 
(i) 
$$ 
\eqalignno{ 
R^\ab - \oot g^\ab R &=  
\la_0 \l[F^{\al\mu} F_\mu^{\ \, \be} - g^\ab L_o (L_\mu) \r] &\cr 
&+ 2 (\pa^\al S) (\pa^\be S) - g^\ab L_o (S)&\cr 
&+ (\hat D^\al \phi)^+ (\hat D^\be \phi) + 
(\hat D^\be \phi)^+ (\hat D^\al \phi) &\cr 
&- g^\ab L_M (\phi) &({\rm III}.13)\cr 
} 
$$ 
(ii) 
$$ 
\la_o D_\mu F^{\mu\al} + \la J^\al = 0  
\eqno({\rm III}.14) 
$$ 
$$ 
J^\al \equiv i \l[ (\hat D^\al \phi)^+ \phi 
- \phi^+ (\hat D^\al \phi) \r] 
\eqno({\rm III}.15) 
$$ 
(iii) 
$$ 
2D_\mu D^\mu S + D_\mu J^\mu = 0 
\eqno({\rm III}.16) 
$$ 
(iv) 
$$ 
(\hat D_\mu \hat D^\mu \phi)^+ + m^2 \phi^+ = 0 
\eqno({\rm III}.17) 
$$ 
(v) 
$$ 
\hat D_\mu \hat D^\mu \phi + m^2 \phi = 0 
\eqno({\rm III}.18) 
$$ 
From equation (III.14) we find 
$$ 
D_\mu J^\mu = 0 
\eqno({\rm III}.19) 
$$ 
The interaction gauge group between the 
electromagnetic field and the matter field 
is given by 
$$ 
\eqalignno{ 
\de_* L_\mu &= \pa_\mu \varphi &({\rm III}.20)\cr 
\de_* \phi &= i \la \varphi \cd \phi &({\rm III}.21)\cr 
\de_* \phi^+ &= -i \la \varphi \cd \phi^+ &({\rm III}.22)\cr 
} 
$$ 
The gauge group for the speedon field is given by 
$$ 
\de_* S = \const 
\eqno({\rm III}.23) 
$$ 
 
\head{IV. Gravity - Electromagnetism - Speedon Field} 
 
Here we study the gravitational interaction in the 
absence of matter, i.e., the Lagrangian (III.3) with $\phi=0$. 
 
We are then left with the Lagrangian 
$$ 
L = \sq{g} R - 2 \la_o \sq{g} L_o (L_\al) -  
2 \sq{g} L_o (S) 
\eqno({\rm IV}.1) 
$$ 
The equations of motion then read 
$$ 
\eqalignno{ 
G^\ab \equiv R^\ab - \oot g^\ab R  
&= \la_o \l[ F^{\al\mu} F_{\mu}^{\ \, \be} - g^\ab L_o (L_\al) \r] 
+ 2 S^\al S^\be - g^\ab L_o (S) &({\rm IV}.2)\cr 
D_\mu F^{\mu\al} &= 0 &({\rm IV}.3)\cr 
D_\al S^\al &= 0 &({\rm IV}.4)\cr 
} 
$$ 
From Appendix B we find the solutions of these equations in  
the case of the static Schwarzschild metric and with the speedon 
field $S$ as the independent variable as  
$$ 
\eqalignno{ 
r &= - a c_1 {\sinh [c_2 S + c_3] \ov \sinh [c_3] \sinh [c_1 S]} 
&({\rm IV}.5)\cr 
e^A &= \l[ {\sinh [c_3] \ov \sinh [c_2 S + c_3]}\r]^2  
&({\rm IV}.6)\cr 
e^B &= {c_1^2 \ov \sinh^2 \l[c_1 S] [c_2\, coth 
[c_2 S + c_3] - c_1\, coth [c_1 S]\r]^2} 
&({\rm IV}.7)\cr 
L&= - {b \ov a} {c_2 \ov \mu^2} 
\l[ coth [c_2 S + c_3] - coth [c_3]\r] 
&({\rm IV}.8)\cr 
} 
$$ 
$a$ and $b$ are integration constants and we have the relations 
$$ 
\eqalignno{ 
\mu^2 &= \oot \la_o \l({b\ov a}\r)^2 &({\rm IV}.9)\cr 
c_1^2 &= 1 + c_2^2 &({\rm IV}.10)\cr 
c^2_2 &= \mu^2 \sinh^2 [c_3] &({\rm IV}.11)\cr 
} 
$$ 
At this time we are only interested in the speedon field.  Turning 
off the electromagnetic interaction means $\la_o = 0$.  This  
implies $c_2 = 0$ and $c_1 = 1$. 
 
The solutions then become 
$$ 
\eqalignno{ 
r &= - {a \ov \sinh [S]} &({\rm IV}.12)\cr 
e^A &= 1 &({\rm IV}.13)\cr 
e^B &= {1 \ov \cosh^2 [S] } &({\rm IV}.14)\cr 
} 
$$ 
or as functions of the radial variable $r$ we get 
$$ 
\eqalignno{ 
S &= - \ln \l[ {a \ov r} + \sq{1 + \l({a \ov r}\r)^2} \r] 
&({\rm IV}.15)\cr 
e^A &= 1 &({\rm IV}.16)\cr 
e^{-B} &= 1 + \l( {a \ov r} \r)^2 &({\rm IV}.17)\cr 
} 
$$ 
This is the solution of gravitational interaction with a massless 
scalar field for the special value $\al=0$; see Appendix C. 
 
From (A.22) we now get an expression for the speed of light 
$$ 
{c^2 \ov c_o^2} = \sq{1 + \l({a \ov r}\r)^2} - {a \ov r}  
\eqno({\rm IV}.18) 
$$ 
or written differently 
$$ 
{c^2 \ov c_o^2} =  
{ {\ts r \ov \ts a} \ov 1 + \sq{1 + \l({\ts r \ov \ts a}\r)^2}} 
\eqno({\rm IV}.19) 
$$ 
As $r \ra 0$ the speed of light goes to zero and as $r \ra \inf$ 
the speed of light increases and reaches the reference speed 
$c_o$. 
 
At $r = a$ one finds $c=0.64\, c_o$.   
 
Observe that as $r$ approaches zero the local physical time becomes 
very large.  There is another interesting observation. 
The concept of gravitational redshift says that the rates of  
similar clocks, located at different places in a gravitational 
field is given by 
$$ 
\l[ {N(2) \ov N(1)} \r]^2 = 
{g_{oo} (2) \ov g_{oo}(1) } 
\eqno({\rm IV}.20) 
$$ 
Let $R$ be a reference radius  where the rate of the clock is $N_R$  
and let $N$ be the rate of a similar clock at the position $r$. 
 
For the exterior Schwarzschild metric one gets then 
$$ 
{N \ov N_R} = 
\l[ {1 - {a \ov r} \ov 1 - {a \ov R} } \r]^{1/2} 
\eqno({\rm IV}.21) 
$$ 
where $a$ is the Schwarzschild radius 
$$ 
a = 2 G_o M 
\eqno({\rm IV}.22) 
$$ 
For $r > R$ and ${a \ov R} \ll 1$, we find 
$$ 
{N \ov N_R} = 1 + \oot [{a \ov R} - {a \ov r} ] 
+ {1 \ov 8}  
\l[ 3 \l({a \ov  R}\r)^2 - 2 \l( {a \ov R} \r) 
\l({a \ov r}\r) - \l( {a \ov r} \r)^2 \r] + \dots 
\eqno({\rm IV}.23) 
$$ 
This applies in particular to similar atomic clocks, one located in Paris and 
one in Boulder; the one in Boulder ticks faster. 
 
For the gravitational field due to the speedon field alone (IV.16) there is  
no gravitational redshift.  We now compare the speed of light $c_R$ at the 
reference radius $R$ with the speed of light $c$ at the position $r$. 
 
For the integration constant $a$ in (IV.18) we take the 
Schwarzschild radius (IV.22). 
 
Then 
$$ 
\eqalignno{ 
c_R &= c_o \l[ \sq{1 + \l( {a \ov R}\r)^2} - {a \ov R} \r]^{1/2} 
&({\rm IV}.24)\cr 
c&= c_o \l[ \sq{1 + \l( {a \ov r}\r)^2} - {a \ov r} \r]^{1/2} 
&({\rm IV}.25)\cr 
} 
$$ 
gives the expansion for $r>R, {a \ov R} \ll 1$ 
$$ 
{c \ov c_R} = 1 + \oot \l[ {a \ov R} - {a \ov r} \r] 
+ {1 \ov 8} \l[ \l({a\ov R}\r)^2 - 2 \l( { a \ov R} \r)  
\l( {a \ov r} \r) + \l({a \ov r} \r)^2 \r] + \dots 
\eqno({\rm IV}.26) 
$$ 
This expression agrees with (IV.25) up to first order.  
 
\vfil\eject 
\head{V. Conclusion} 
 
In electrodynamics, elimination of the physical time through 
dimensional analysis, identifies the speed of light as a dilaton field. 
In the Lagrange formalism for gravitational interactions with the 
electromagnetic field and a charged matterfield this point of view 
introduces a massless scalarfield $S$ that we call the speedon field. 
In the absence of the matter field we found solutions for the 
gravitational-electromagnetic-speedon field equations.  If 
we now turn off the electromagnetic interaction we are left with 
the speedon field.  This is a particular solution of the  
gravitational interaction with a massless scalar field.  The pure 
speedon field gives an expression for the speed of light.  For 
large values of the radial coordinate the speed of light 
becomes constant and for small values of the radial coordinate the speed 
of light goes to zero.  This then raises the question about the meaning of a  
local physical time.  The analytic solution in the presence of a 
matterfield should give some more information.  According to Vandyck [2] very accurate  
redshift magnitude curves should  provide information about the presence 
of a dilaton field.  It is interesting to observe that gravitatoinal interaction 
as treated in this paper introduces an ``index of refraction" for 
the Universe. 
 
\vfil\eject 
 
\head{Appendix A: Scaling the Electromagnetic Field} 
 
Electrodynamics is described by a vector field $\{ A_\mu \}$, 
the so-called vectorpotential, which belongs to the restmass zero, 
spin one representation of the Lorentz group. 
 
The physical dimension of $A_\mu$ is 
$$ 
[A_\mu ] = M L^2 T^{-2} Q^{-1} 
\eqno({\rm A}.1) 
$$ 
with $c$ denoting the speed of light, $t$ the physical time and 
$$ 
\{ A_\mu\} = (A_o, \vecA) 
\eqno({\rm A}.2) 
$$ 
the electric field is given by 
$$ 
\vecE = {1 \ov c} \pp {\vecA} t - {\rm grad}\ A_o 
\eqno({\rm A}.3) 
$$ 
and the magnetic field by 
$$ 
\vecB = - {1 \ov c} {\rm curl}\ \vecA  
\eqno({\rm A}.4) 
$$ 
These fields have the physical dimensions 
$$ 
\eqalignno{ 
[E] &= MLT^{-2} Q^{-1} &({\rm A}.5)\cr 
[B] &= MT^{-1} Q^{-1} &({\rm A}.6)\cr 
} 
$$ 
We now introduce the geometric time $x^o$, which is 
related to the physical time $t$ by 
$$ 
x^o = ct 
\eqno({\rm A}.7) 
$$ 
and the scaled field $L_\mu$ through 
$$ 
L_\mu = {1 \ov c^2} A_\mu. 
\eqno{\rm A}.8) 
$$ 
$L_\mu$ has the physical dimension 
$$ 
[L_\mu] = M Q^{-1} 
\eqno({\rm A}.9) 
$$ 
which does not depend on the physical time.  Maxwell's equations follow 
from a variational principle with the Lagrangian 
$$ 
L_o (L_\mu) = {1 \ov 4\pi K_o} \cd {1 \ov 4} 
F_\mn F^\nm 
\eqno({\rm A}.10) 
$$ 
where 
$$ 
K_o = {K \ov c^2} , 
\eqno({\rm A}.11) 
$$ 
with $K$ being the Coulomb constant and  
$$ 
F_\mn \equiv \pa_\mu L_\nu - \pa_\nu L_\mu 
\eqno({\rm A}.12) 
$$ 
The physical dimensions are  
$$ 
\eqalignno{ 
[K_o] &= MLQ^{-2} &({\rm A}.13)\cr 
[L_o] &= ML^{-3} &({\rm A}.14)\cr 
} 
$$ 
The Lagrangian $L_o$ is a mass density.   
 
Through scaling with the help of the speed of  
light electrodynamics is thus formulated in such  a way 
that it does not depend on physical time.  This now opens 
the possibility that the speed of light could be a function 
on Minkowski space [1].  With 
$$  
\{ L_\mu \} = (L_o, \vecL)  
\eqno({\rm A}.15) 
$$ 
and $c$ being the local speed of light, we can now retrieve 
the local quantities 
$$ 
\eqalignno{ 
A_\mu &= c^2 L_\mu &({\rm A}.16)\cr 
\vecE &= c^2 \l[ \pp {\vecL} {x^o} - {\rm grad}\ L_o \r] 
&({\rm A}.17)\cr 
\vecB &= - c \ {\rm curl}\ \vecL &({\rm A}.18)\cr 
t &= {1 \ov c} x^o &({\rm A}.19)\cr 
} 
$$ 
The physical time thus becomes a local concept. 
 
The scaling can be written as 
$$ 
L_\mu = {c_o^2 \ov c^2} {1 \ov c_o^2} A_\mu 
\eqno({\rm A}.20) 
$$ 
with $c_o$ as a reference speed of light.  Then 
$$ 
L_\mu = e^{-S} {1 \ov c_o^2} A_\mu 
\eqno({\rm A}.21) 
$$ 
where  
$$ 
S = \ln \l( {c^2 \ov c_o^2} \r) 
\eqno({\rm A}.12) 
$$ 
$S$ is thus a dilation field [2] and belongs to the restmass 
zero, spin zero representation of the Loretnz group.  Since  
$S$ is related to the speed of light we call it the speedon field. 
 
\bigskip 
 
\head{Appendix B:  The Schwarzschild metric and the solutions} 
 
The Schwarzschild metric is given by 
$$ 
ds^2 = e^A (dx^o)^2 - e^B (dr)^2 
- r^2 \{ d \th^2 + \sin^2 \th d \vp^2 \} 
\eqno({\rm B}.1) 
$$ 
For the static case $A= A(r), B=B(r).$  With 
$g= - {\rm Det} (g_\ab)$ we find 
$$ 
\sq{g} = e^{\oot (A+B)} r^2 \sin \th 
\eqno({\rm B}.2) 
$$ 
and the Einstein tensor [3] 
$$ 
G^\al_{\ \, \be} = g^{\al\ga}  
[R_{\ga\be} - \oot g_{\ga\be} \cd R] 
\eqno({\rm B}.3) 
$$ 
reads 
$$ 
\eqalignno{ 
G^0_{\  0} &= - {1 \ov r^2} {d \ov dr} 
\l[ r (e^{-B} - 1) \r] &({\rm B}.4)\cr 
G^1_{\ 1} &= G^0_{\ 0} - {1 \ov r} e^{-B} 
{d \ov dr} (A+B) &({\rm B}.5)\cr 
G^2_{\ 2} &= {1 \ov 2r} {d \ov dr} 
\l[r^2 G^1_{\ 1} \r] - {1 \ov 4} e^{-B} 
\dd A r {d \ov dr} (A+B)  
&({\rm B}.6)\cr 
G^3_{\ 3} &= G^2_{\ 2} &({\rm B}.7)\cr 
} 
$$ 
All other components vanish. 
 
We now introduce the auxiliary functions 
$$ 
\eqalignno{ 
f &= e^{\oot (A+B)} &({\rm B}.8)\cr 
h&= r e^{\oot (A-B)} &({\rm B}.9)\cr 
} 
$$ 
Then 
$$ 
\eqalignno{ 
e^A &= {1 \ov r} fh &({\rm B}.10)\cr 
e^B &= {rf \ov h} &({\rm B}.11)\cr 
} 
$$ 
and 
$$ 
\sq{g} = r^2 f \sin \th  
\eqno({\rm B}.12) 
$$ 
The relevant components of the Einstein tensor then become 
$$ 
\eqalignno{ 
G^0_{\ 0} &= - {1 \ov r^2} {d \ov dr} 
\l[ {h \ov f} - r \r] &({\rm B}.13)\cr 
G^1_{\ 1} &= G^0_{\ 0} - 2 {h \dd f r \ov r^2 f^2} 
&({\rm B}.14)\cr 
G^2_{\ 2} &= {1 \ov 2r} {d \ov dr} [r^2 G^1_{\ 1} ]  
- \oot {h \dd f r \ov r f^2} {d \ov dr} \ln  
\l( {fh \ov r} \r) &({\rm B}.15)\cr 
} 
$$ 
With $(L_\al) = (L, 0, 0, 0)$ we find the Lagrangian 
for the electromagnetic field as 
$$ 
L_0 (L_\al) = {1 \ov 2f^2} \dd L r \dd L r  
\eqno({\rm B}.16) 
$$ 
For the speedon field the Lagrangian reads 
$$ 
L_o (S) = - { h \ov rf} \dd S r \dd S r  
\eqno({\rm B}.17) 
$$ 
The equations of motion then become 
$$ 
\eqalignno{ 
G^0_{\ 0} &= \oot \la_o {1 \ov f^2} \dd L r \dd L r 
+ {h \ov rf} \dd S r \dd S r &({\rm B}.18)\cr 
G^1_{\ 1} &= \oot \la_o {1 \ov f^2} \dd L r \dd L r 
- {h \ov rf} \dd S r \dd S r &({\rm B}.19)\cr 
G^2_{\ 2} &= - \oot \la_o {1 \ov f^2} \dd L r \dd L r 
+ {h \ov rf} \dd S r \dd S r &({\rm B}.20)\cr 
} 
$$ 
$$ 
\eqalignno{ 
{d \ov dr}\l[ {r^2 \ov f} \dd  L r \r] &= 0 &({\rm B}.21)\cr 
{d \ov dr}  \l[ rh \dd S r \r] &= 0 
&({\rm B}.22)\cr 
} 
$$ 
We now find the following system of independent equations 
$$ 
\eqalignno{ 
\dd L r &= b {f \ov r^2} &({\rm B}.23)\cr 
\dd S r &= {a \ov rh} &({\rm B}.24)\cr 
{1 \ov f} \dd f r &= r \dd S r \dd S r &({\rm B}.25)\cr 
\dd h r &= f \l[ 1 - \oot \la_o b^2 {1 \ov  r^2} \r] 
&({\rm B}.26)\cr 
} 
$$ 
Introducing the dimensionless quantities 
$$ 
r = a x \ \ , \ \ h = ak 
\eqno({\rm B}.27) 
$$ 
and the abbreviation $' \equiv {d \ov dx}$, we find the 
equation of motion to read 
$$ 
\eqalignno{ 
L' &= {b \ov a} {1 \ov x^2} f &({\rm B}.28)\cr 
S' &= {1 \ov xk} &({\rm B}.29)\cr 
{f' \ov f} &= x S' S' &({\rm B}.30)\cr 
k' &= f 
\l[ 1 - \oot \la_o \l( {b \ov a} \r)^2  \cd {1 \ov x^2} \r] 
&({\rm B}.31)\cr 
} 
$$ 
As in [9, 10] we look upon the speedon field as the independent 
variable and denote by $\cd  \equiv {d \ov d S}$.  The equations of motion 
then read 
$$ 
\eqalignno{ 
\dot L &= {b \ov a} {\dot x \ov x^2} f &({\rm B}.32)\cr 
\dot x \dot f &= xf &({\rm B}.33)\cr 
\dot x &= xk &({\rm B}.34)\cr 
\dot k &= \dot x f \l[ 1 - \mu^2 {1 \ov x^2} \r] &({\rm B}.35)\cr 
} 
$$ 
with 
$$ 
\mu^2 = \oot \la_o \l( {b \ov a} \r)^2. 
$$ 
The boundary conditions we impose are 
$$ 
S \ra 0 \quad \Ra f \ra 1,  x \ra \inf,  
{k \ov x} \ra 1, L \ra 0,  
\eqno({\rm B}.36) 
$$ 
implying asumptotic flatness, and the vanishing of the speedon 
field at infinity. 
 
To solve these equations we introduce the auxiliary function $F(S)$ through 
$$ 
{\dot x \ov x} = F 
\eqno({\rm B}.37) 
$$ 
Then from (B.33) we get 
$$ 
{ \dot f \ov f} = {1 \ov F} 
\eqno({\rm B}.38) 
$$ 
This leads to 
$$ 
\eqalignno{ 
x &= x_o e^{\int FdS} &({\rm B}.39)\cr 
f &= f_o e^{\int {1 \ov F} d S} &({\rm B}.40)\cr 
} 
$$ 
Equations (B.34) and (B.35) then become 
$$ 
k = F 
\eqno({\rm B}.41) 
$$ 
$$ 
\dot F = F\, x_o f_o e^{\int \l[ F + {1 \ov F} \r] dS} 
\l[ 1 - \mu^2 {1 \ov x_o^2} e^{-2 \int FdS} \r] 
\eqno({\rm B}.42) 
$$ 
With  
$$ 
\al \equiv x_o f_o, \ \ \be = \mu^2 f_o {1 \ov x_o} 
\eqno({\rm B}.43) 
$$ 
equation (B.42) reads 
$$ 
{\dot F \ov F} = \al e^{\int \l[ {1 \ov F} + F \r] dS} 
- \be e^{\int \l( {1 \ov F} - F \r) dS} 
\eqno({\rm B}.44) 
$$ 
Differentiating this equation and substituting it we get 
$$ 
{d \ov dS} 
\l[ {\dot F \ov F} \r] = { \dot F \ov F^2} 
+ F \l\{ \al e^{\int \l[ {1 \ov F} + F \r] dS} 
+ \be e^{\int \l[ {1 \ov F} - F \r] dS} 
\r\} 
\eqno({\rm B}.45) 
$$ 
Adding and subtracting equations (B.44) and (B.45) and 
using the abbreviations 
$$ 
\eqalignno{ 
X &= { \dot F \ov F} + F + {1 \ov F} &({\rm B}.46)\cr 
Y &= { \dot F \ov F} - F + {1 \ov F} &({\rm B}.47)\cr 
} 
$$ 
results in  
$$ 
\eqalignno{ 
2 \al e^{\int \l[ {1 \ov F} + F \r] dS} &= {1 \ov F} \dot X &({\rm B}.48)\cr 
2 \be e^{\int \l[ {1 \ov F} - F \r] dS} &= {1 \ov F} \dot Y &({\rm B}.49)\cr 
} 
$$ 
and  
$$ 
\eqalignno{ 
2F &= X - Y &({\rm B}.50)\cr 
2 \l[ { \dot F \ov F}  + {1 \ov F}\r] &= X + Y &({\rm B}.51)\cr 
} 
$$ 
Taking the logarithmic derivative of equations (B.48) and  
(B.49) we get 
$$ 
\eqalignno{ 
{1 \ov F} + F &= -  {\dot F \ov F} + {\ddot X \ov \dot X} 
&({\rm B}.52)\cr 
{1 \ov F} - F &= -  {\dot F \ov F} + {\ddot Y \ov \dot Y} 
&({\rm B}.53)\cr 
} 
$$ 
resulting in the equations 
$$ 
\eqalignno{ 
\ddot X &= X \dot X &({\rm B}.54)\cr 
\ddot Y &= Y \dot Y &({\rm B}.55)\cr 
} 
$$ 
Observe that from equations (B.10) and (B.11) 
we have the relations 
$$ 
\eqalignno{ 
\dot X &= 2 F^2 e^B &({\rm B}.56)\cr 
\dot Y &= 2 \mu^2 e^A &({\rm B}.57)\cr 
} 
$$ 
Asymptotic flatness then demands 
$$ 
\dot Y (0) = 2 \mu^2 \eqno({\rm B}.58) 
$$ 
and that both $\dot X $ and $\dot Y$ are positive. 
 
The solutions of (B.54) and (B.55) that are compatible 
with the boundary conditions (B.36), (B.58) are given by 
$$ 
\eqalignno{ 
X &= - 2 c_1\, coth [ c_1 S ] &({\rm B}.59)\cr 
Y &= - 2 c_2\, coth [ c_2 S + c_3] &({\rm B}.60)\cr 
} 
$$ 
From (B.50) and (B.51) we then get 
$$ 
F = c_2\, coth [c_2 S + c_3] - c_1\, 
coth [c_1 S] 
\eqno({\rm B}.61) 
$$ 
and the condition 
$$ 
c^2_1 = 1 + c^2_2 
\eqno({\rm B}.62) 
$$ 
The condition (B.58) then becomes 
$$ 
c^2_2 = \mu^2 \sinh^2 [c_3] 
\eqno({\rm B}.63) 
$$ 
We then find the solutions of (B.32) to (B.35) 
$$ 
\eqalignno{ 
x &= - {c_1 \ov \sinh [c_3]} 
{\sinh [c_2 S + c_3] \ov \sinh [c_1 S]} 
&({\rm B}.64)\cr 
f &= {c_1 \sinh  [c_3] \ov \sinh [c_2 S + c_3] \sinh [c_1 S] 
\{ c_1\, coth [c_1 S] - c_2\, coth [c_2 S + c_3] \} } &({\rm B}.65)\cr 
k &= c_2\, coth [c_2 S + c_3] - c_1\, coth [c_1 S] &({\rm B}.66)\cr 
L &= - {b \ov a} {c_2 \ov \mu^2} 
\l[ coth [ c_2 S + c_3] - coth [c_3] \r] &({\rm B}.67)\cr 
} 
$$ 
From (B.56) and (B.57) we get 
$$ 
\eqalignno{ 
e^A &= \l[ {\sinh [c_3] \ov \sinh [c_2 S + c_3]} \r]^2 
&({\rm B}.68)\cr 
e^B &= {c_1^2 \ov \sinh^2 [c_1 S] [c_2\,coth [c_2 S + c_3] - c_1\, coth [c_1S] ]^2} 
&({\rm B}.69)\cr 
} 
$$ 
 
\bigskip 
 
\goodbreak 
\head{Appendix C:  The Massless Scalarfield} 
 
Gravitational interaction with a massless scalarfield alone is given 
by setting $\la_o = 0$ in (IV.1).  This corresponds to 
setting $\mu = 0$ in (B.57).  The equations (B.59) and (B.60) 
then are replaced by 
$$ 
\eqalignno{ 
X&= - 2c_1\,coth [c_1 S] &({\rm C}.1)\cr 
Y&= 2\al &({\rm C}.2)\cr 
} 
$$ 
where $\al$ is a constant. 
 
From (B.50) and (B.51) we then get 
$$ 
F = - \al - c_1\, coth [c_1 S] 
\eqno({\rm C}.3) 
$$ 
and the condition 
$$ 
c_1^2 = 1 + \al^2  
\eqno({\rm C}.4) 
$$ 
The complete solution [4] is then given by 
$$ 
\eqalignno{ 
x &= - c_1 e^{-\al S} {1 \ov \sinh [c_1 S] } &\cr 
f &= c_1 e^{\al S} {1 \ov c_1 \cosh [c_1 S] + \al \sinh [c_1 S] } 
&({\rm C}.5)\cr 
k &= - \al - c_1\,coth [c_1 S] &({\rm C}.6)\cr 
e^A &= e^{2 \al S} &({\rm C}.7)\cr 
e^B &= \l[{c_1 \ov c_1 \cosh [c_1 S] + \al \sinh [c_1 S] } \r]^2  
&({\rm C}.8)\cr 
} 
$$

\def\pa{{\it Physica A} } 
\bigskip 
 
\head{References}

\no 1 W. Wyss, \fpl {\bf 6} (1993), 591. 
 
\no 2 M. A. Vandyck, {\it Class. Quantum Grav.} {\bf 12} (1995), 209. 
 
\no 3 M. Carmeli, ``Classical Fields: General Relativity and Gauge 
Theory," John Wiley and Sons, 1982. 
 
\no 4 W. Wyss, ``The Energy-Momentum Tensor for Gravitational 
Interactions" (to be published). 
 
\no 5 W. Wyss, ``The Energy-Momentum Tensor in Classical Field 
Theory" (to be published). 
 
\no 6 W. Wyss, ``The Relation between the Gravitational Stress 
Tensor and the Energy-Momentum Tensor" (to be published in 
{\it Helv. Phys. Acta}). 
 
\no 7 C. W. Misner, K. S. Thorne, J. A. Wheeler, ``Gravitation," 
W. H. Freeman and Company, 1973, p. 504. 
 
\no 8 Frank E. Schroeck, Jr., ``Quantum Mechanics on Phase Space," 
Kluwer Academic Publishers, 1996, p. 453. 
 
\no 9 M. Wyman, \prd {\bf 24} (1981), 839. 
 
\no {10} K. Schmoltzi, Th. Schucker, \pla {\bf 161} (1991), 212.

\bye